\documentstyle[aps]{revtex}
\begin{document}
\title{Non universal diffusion limited aggregation and exact fractal
dimensions}
\author{P. Ossadnik}
\address{HLRZ, KFA J\"ulich, Postfach 1913, 52425 J\"ulich, Germany
\footnote{Present address: Center for Polymer Studies and Department of
Physics, Boston University, 590 Commonwealth Ave., Boston,~MA~02215, USA}}
\author{Chi-Hang Lam}
\address{Department of Applied Physics, Yale University, Box 208283,
New Haven, CT 06520-8283, USA.}
\author{Leonard M. Sander}
\address{H.M.~Randall~Laboratory of Physics,
The University of Michigan, Ann Arbor, MI 48109-1120, USA}

\maketitle

\begin{abstract}
In analogy to recent results on non-universal roughening in surface growth [Lam
and Sander, Phys. Rev. Lett. {\bf 69}, 3338 (1992)], we propose a variant of
diffusion-limited aggregation ($DLA$) in which the radii of the particles are
chosen from a power law distribution. For very broad distributions, the huge
particles dominate and the fractal dimension is calculated exactly using a
scaling theory. For narrower distributions, it crosses back to DLA.  We
simulated $1200$ clusters containing up to $200,000$ particles. The fractal
dimensions obtained are in reasonable agreement with our theory.
\end{abstract}

\pacs{PACS  numbers: }

\draft
%\psdraft
\narrowtext

Diffusion limited aggregation ($DLA$) \cite{Witten} is an important prototype
of fractal growth\cite{Vicsekbook}. In two dimensions, its fractal dimension is
determined to be $ 1.712 \pm 0.003$ from large scale off-lattice
simulations\cite{Meakin,Ossadnik}. There exists no exact formula for the
fractal dimension. In spite of intensive studies, the task of obtaining
accurate estimates of the fractal dimension by analytical means remains to be
very challenging\cite{Vicsekbook,Halsey}.

In this work, we investigate a new class of variants with $exactly$ solvable
fractal dimensions for a certain range of a model parameter.  In another range,
the model crosses back to $DLA$.  The transition is associated with an
interesting morphology change.  The model might have relevance to
cluster-cluster aggregation ($CCA$) models which simulate colloidal
aggregation\cite{Vicsekbook}.  More importantly, $DLA$ and most previous
non-trivial variants are not expected to have exactly solvable fractal
dimensions.  Our variant might have a particular role in testing theoretical
approaches of branched growth\cite{Vicsekbook,Halsey}.  However, even though
the variant has an exact dimension for a certain regime which extrapolates
continuously to $DLA$, our approach is not capable of calculating the fractal
dimension, ${D_{DLA}}$, of $DLA$. The reason is that the transition point is
not known a priori and has to be expressed in terms of ${D_{DLA}}$.  We
concentrate on the two dimensional case, while the results can be generalized
easily to higher dimensions.

Our variant of $DLA$ is motivated by recent results on Zhang's model of surface
growth with power-law noise \cite{Zhang,Lam}.  The algorithm is very similar to
standard $DLA$. Particles are launched one by one and carry out Brownian
motion. Starting from an immobile seed, the aggregate grows when a walker hits
it and becomes part of it.  Usually, all the particles launched have the same
radii.  The asymptotic scaling properties and the morphology are not expected
to change if the particle radii are random but bounded.  However, the
universality can be broken if the probability distribution of the radii is very
broad.  This leads us to characterize our variant by the following power-law
distribution, $P(r)$, of particle radii, $r$:
\begin{equation}
\label{rdistribution}
P(r)= \left\{ \begin{array}{lll}
     {\mu}/{r^{\mu+1}} &~~~~~~~~& \mbox{for~} r \ge 1\\
	0              &	& \mbox{for~}r<1
\end{array} \right.
\end{equation}
It recovers the standard $DLA$ algorithm as $\mu \rightarrow \infty$.  In
general, the walkers can be fractals themselves with dimension $D_p$.  An
interesting example is the $CCA$ model in which, for some regimes, the radii of
the ensemble of clusters can have a power-law distribution\cite{Vicsekbook}.
The walkers in our variant then correspond to the wandering clusters in the
$CCA$ model. In this case, the fractal dimension of the walkers and the final
aggregate should be the same.  This property follows naturally from our scaling
theory in the appropriate regime.

We focus on the case that the aggregate and the individual particles are dense
enough so that they are not transparent to each other. This is generally true
in low dimensions.  For programming convenience, we further assume in our
simulations that the wandering particles have a circular outer boundary. This
simplification should not alter the scaling properties.

The simulation of off-lattice $DLA$ with the power law distribution of the
particle radii in Eq. (\ref{rdistribution}) is complicated due to the absence
of an upper cutoff for the particle radii. To determine accurately whether a
walker has touched the cluster, it is no longer sufficient to search in a
finite neighborhood for centers of other particles. In order to obtain an
efficient algorithm, the focus must be shifted from the centers of the
particles to their perimeters, which are instead stored in a hierarchical map
structure\cite{Meakin,Ossadnik}. We can thus check efficiently whether a walker
with a given radius overlaps with a perimeter site of some particle on the
cluster. However, this procedure significantly increases both the storage
requirements and the run time of the simulation, especially for a small power
law exponent $\mu\rightarrow 1$. Therefore, we had to restrict our simulations
to clusters with masses between $100,000$ and $200,000$ particles.

Figure \ref{clusters} shows three typical clusters grown for exponents
$\mu=2.5$, $\mu=1.713$ and $\mu=1.3$. Each of them contains $100,000$
particles.  The choice of the values of $\mu$ is to illustrate a morphology
transition which will be explained later using a scaling theory. For $\mu=2.5$
[Fig.
\ref{clusters}(a)], the presence of the large particles results in a cluster
significantly more noisy than $DLA$. However, the small particles are abundant
enough to dominate the geometry and the overall branching structure is very
similar to $DLA$. There exists a number of well defined main branches emanating
radially from the center of the cluster. We observe that the ratio between the
size of the largest particles and that of the cluster decreases as the number
of particles, $N$, increases. In fact, we will explain that the cluster becomes
indistinguishable to $DLA$ for very large $N$. In contrast, for the very broad
distribution at $\mu=1.3$ [Fig.
\ref{clusters}(c)], the morphology is completely altered by the very big
particles. A well defined geometrical center of the cluster does no longer
exist.  The size of the largest particles is comparable to that of the cluster
and dominates the structure completely. The morphology is more similar to that
of $CCA$ than $DLA$. We will show that the transition between the two different
morphologies occurs at $\mu={D_{DLA}}\simeq 1.713$ [Fig.
\ref{clusters}(b)].  Here, the $DLA$ type morphology holds marginally
and the size of the largest particles compared to the cluster size
decreases very slowly as $N$ increases.

We now present a scaling theory for the variant, which predicts a transition
from a regime dominated by small particles to one with dominating large ones as
$\mu$ decreases.  Consider the scaling form
\begin{equation}
\label{RN}
R_G \sim N ^ {1/\gamma}
\end{equation}
where $R_G$ is the radius of gyration of the cluster and $N$ is the number of
particles. For cases such as $D_p=0$ or $\mu\rightarrow \infty$, the exponent
$\gamma$ reduces to the fractal dimension. The general relationship will be
worked out later.

We first examine a broad distribution with $\mu<{D_{DLA}}$. We are going to
show that $\gamma$ is $exactly$ given by $\gamma=\mu<{D_{DLA}}$. The idea of
the proof to be presented below can be sketched out briefly as follows: If it
were true that $\gamma>\mu$, the cluster radius $R_G$ would be even smaller
than the typical size of the largest particle inside the cluster. If
$\gamma<\mu$, $R_G$ would be too large to be accounted for by the large
particles. This leads to a domination of small particles, which again can be
proved to be wrong.  Our derivation is closely related to the analogous one for
surface growth with power-law noise, in which case Lam and Sander
\cite{Lam} proved the exactness of a formula for the scaling exponent
suggested independently by Zhang \cite{ZhangB} and Krug \cite{Krug}.

We now give our arguments in more detail.  It is easy to show that $\gamma \ne
{D_{DLA}}$, which necessarily implies a non-universal behavior $\mu<{D_{DLA}}$.
For a cluster of $N$ particles, the expected radius of the largest particle,
$r_{max}$, sampled from the power-law distribution in Eq.
(\ref{rdistribution}), follows $r_{max}\sim N^{1/\mu}$. The proposition
$\gamma={D_{DLA}}$ implies $R_G\sim N^{1/{D_{DLA}}}$.  This is impossible for
$\mu<{D_{DLA}}$ because at sufficiently large $N$, the largest particle would
even be bigger than the cluster itself ($r_{max}> R_G$)! The same argument
shows more generally that $\gamma>\mu$ is false when $\mu<{D_{DLA}}$.

In fact, the huge particles are not only relevant to the scaling, as proved
above, but also dominate the geometry completely. It is plausible that the
largest particle sets the scale of the cluster: $R_G \sim r_{max}$.  This
assumption immediately leads to $\gamma=\mu$, since $r_{max}\sim N^{1/\mu}$. We
will show that this simple picture is indeed correct asymptotically.

The conventional way to compare the visual appearance of $DLA$ of
different sizes is to rescale them to a standardized radius of
gyration. Instead, we now rescale the clusters by a factor
$N^{-1/\mu}$ and examine the dependence of the radii of the clusters
on $N$. In this rescaling scheme, the size of the biggest particle is
independent of $N$. Furthermore, the radii, $r'$, of the rescaled
constituent particles follow a new probability distribution $P'(r')$.
In a cluster of $N$ particles, the number density of particles per
unit $r'$ is $n(r')=NP'(r')$ given by
\begin{equation}
\label{n}
n(r')= \left\{ \begin{array}{lll}
{\mu}/{r'^{\mu+1}} &~~~~~~~~& \mbox{for~} r' \ge r_m\\
	0	  &        & \mbox{for~} r'<r_m
\end{array} \right.
\end{equation}
where the lower cutoff radius, $r_m$, is a function of $N$ given by
$r_m(N) = N^{-1/\mu}$. This implies that not only $r_{max}$ but also
the whole distribution $n(r')$, except for the lower cutoff $r_m$, is
independent of $N$. The simple picture of the domination of the large
particles corresponds to the assumption that the variations of the
cutoff can be neglected asymptotically. If this is true, $n(r')$ is
completely independent of $N$ as $N \rightarrow \infty$ and both, the
self-similarity and $\gamma=\mu$, become obvious.

%-------- can one change this (or even take it out?)
%
%The effect of the variation in the cutoff can be accessed as follows.  When
%%$N$
%increases from $N_1$ to $N_2$, the cluster differs by having additional small
%particles with radii $r'$ in the range $[r_m(N_2),r_m(N_1)]$.  They obviously
%increase the cluster size.  However, whether they lead to an increase in the
%asymptotic exponent $1/\gamma$ is not trivial. Suppose the exponent did admit
%corrections so that $\gamma < \mu$. The rescaled cluster has a radius $R'\sim
%N^{1/\gamma-1/\mu}$. For $r_m$ sufficiently close to zero, which corresponds
%%to
%large $N$, $R'$ can be arbitrarily large. In this case, the very small
%particles are responsible for all the space occupied by the cluster and they
%dominate the geometry completely. Universality with standard $DLA$ should
%%hold.
%However, this is proved to be false for $\mu<{D_{DLA}}$. Hence, $R'$ has to
%%converge
%to some finite value as $N$ increases and $\gamma=\mu$ follows.
%
%-------- can one change this (or even take it out?)

For narrower distributions with $\mu>{D_{DLA}}$, the small particles dominate.
We have $\gamma={D_{DLA}}$ and the morphology is the same as $DLA$.  The huge
particles are irrelevant to the geometry. The radius of the largest particle
vanishes compared with the cluster size as $N$ increases. In summary, we have:
\begin{equation}
\label{gamma}
\gamma = \min \{ \mu, {D_{DLA}} \}
\end{equation}

Although the complication about the lower cutoff does not alter the exponent,
it leads to strong crossover effects. Using the above rescaling scheme with the
factor $N^{-1/\mu}$, as $N \rightarrow \infty$ the rescaled cluster radius,
$R'$, converges for $\mu<{D_{DLA}}$ but diverges for $\mu>{D_{DLA}}$. The
$\mu={D_{DLA}}$ case is marginal and $R'$ is expected to diverge
logarithmically, and by analogy with the results for surface growth\cite{Lam}
we postulate a logarithmic correction to scaling at $\mu={D_{DLA}}$:
\begin{equation}
\label{Rcorrection}
R_G \sim N^{1/{D_{DLA}}} [log(N)]^{1/2}
\end{equation}

Now, we compute the fractal dimension $D$ of the cluster from the relation $M
\sim R_G^D$, where $M$ is the mass of the aggregate.  Using the
distribution of the particle radii in Eq. (\ref{rdistribution}) and the
assumption that the constituent particles are fractals of dimension $D_p$, we
get
\begin{equation}
\label{mass}
M \sim \left\{ \begin{array}{ll}
	N	 & \mbox{~~~ $D_p ~<~ \mu$} \\
        N^{D_p/\mu} & \mbox{~~~ $\mu ~<~ D_p$}
	\end{array}
\right.
\end{equation}
where the averaged particle mass diverges for small $\mu$.  At $\mu = D_p$, the
divergence is marginal:
\begin{equation}
\label{Mcorrection}
M\sim N\ln N
\end{equation}
Combining Eqs. (\ref{RN}), (\ref{gamma}) and (\ref{mass}), the fractal
dimensions are obtained for four different cases:
\begin{equation}
\label{D}
D =  \left\{ \begin{array}{ll}
	{D_{DLA}}	 & \mbox{~~~ $D_p,{D_{DLA}} ~<~ \mu$} \\
        D_p{D_{DLA}}/\mu & \mbox{~~~ ${D_{DLA}} ~<~ \mu ~<~ D_p$} \\
        \mu & \mbox{~~~ $D_p ~<~ \mu ~<~ {D_{DLA}}$} \\
        D_p & \mbox{~~~ $\mu ~<~ D_p,{D_{DLA}}$}
	\end{array}
\right.
\end{equation}
The variant crosses back to $DLA$ for narrower distributions of $D_p,{D_{DLA}}
<
\mu$. For ${D_{DLA}} < \mu < D_p$, even though the morphology is the same as
$DLA$, the fractal dimension is different since the averaged particle mass
diverges.  For the other two cases of broad distributions, the morphology is
non-universal. It is particularly interesting that for $\mu < D_p,{D_{DLA}}$,
$D=D_p$ follows.  It is precisely how it should be when the walker represents
the clusters in the $CCA$ model. This sets an upper bound for $\mu$ if the
variant describes some regime of $CCA$.  Simulations of $CCA$ in general give
$\mu$ or its effective value well within this bound \cite{Vicsekbook}.

We test our scaling theory numerically for the case $D_p=2$.  Equation
(\ref{D}) then reduces to
\begin{equation}
\label{D2}
D \sim  \left\{ \begin{array}{ll}
	{D_{DLA}}	 & \mbox{~~~ $2 ~<~ \mu$} \\
        2{D_{DLA}}/\mu & \mbox{~~~ ${D_{DLA}} ~<~ \mu ~<~ 2$} \\
        2 & \mbox{~~~ $\mu ~<~ {D_{DLA}}$}
\end{array}
\right.
\end{equation}
The verification of Eq.  (\ref{D2}) also establishes the validity of Eq.
(\ref{gamma}) and the general result in Eq. (\ref{D}), since they are all
related by the trivial Eq. (\ref{mass}).  The logarithmic corrections in Eqs.
(\ref{Rcorrection}) and (\ref{Mcorrection}) at the transition points give
respectively:
\begin{equation}
\label{logcorr}
M \sim  \left\{ \begin{array}{ll}
        R_G^{{D_{DLA}}} /log(N) & \mbox{~~~ $\mu = {D_{DLA}}$} \\
	R_G^{2} log(N)& \mbox{~~~ $  \mu = 2$}
\end{array}
\right.
\end{equation}
In our naive computation of $D$ from the slope of the best fitted straight line
in the log-log plot of $M$ against $R_G$, these corrections cause an
underestimation at $\mu = {D_{DLA}}$ and an overestimation at $\mu=2$.

Let $r_i$ and $\vec x_i$ be respectively the radius and the position of the
$i$-th particle in a cluster with $N$ particles.  Assuming that every particle
has uniform unit density, the mass of the individual particle is $m_i=\pi
r_i^2$.  Since the particles are disks instead of points, the mass and radius
of gyration of the cluster is given by:
\begin{eqnarray}
M &=& \sum_{i=1}^{N} m_i \\
M R_G^2 &=& \sum_{i=1}^{N} m_i [( \vec x_i - \vec x_{CM}) ^2+ \frac{1}{2}
r_i^2]
\end{eqnarray}
where $\vec x_{CM}$ is the center of mass of the cluster.

We have grown a total of $1200$ clusters for exponents $\mu$ in the range
$1.0\leq\mu\leq 5.0$. For various cluster sizes $N$, we compute the ensemble
averaged cluster mass and radius of gyration. Each data point in the $R_G$ vs
$M$ plot is obtained from ensemble average over clusters of fixed $N$. However,
the averaging has to be done with caution to avoid divergence. For example,
Eq.  (\ref{rdistribution}) implies that the probability distribution, $P(m)$,
of the mass, $m$, of the walkers follows the power-law: $P(m)\sim
1/m^{\mu/2+1}$. For $\mu<2$, the arithmetic mean of $m$ diverges.
% Thus, the cluster mass $M$ follows the asymmetric Levy stable distribution
% \cite{Gnedenko}, which have the same power-law tail and diverging mean.
% In contrast, the distribution of the variable $\ln M$ has an exponential
% tail.
However, the geometrical ensemble average, $\exp < \ln M>$ is well defined,
where the bracket denotes the arithmetic mean.  Similarly, we also take the
geometrical average $\exp < \ln R_G >$ for the radius of gyration $R_G$.
Figure (\ref{mvsrg}) shows the geometrical ensemble averages $M$ against $R_G$
in a log-log plot for the selected values $\mu = 1.3$, $\mu = 1.7 $, and $\mu =
2.5$. The numbers of clusters used are 50, 104 and 25 respectively.  For all
values of $\mu$ we investigated, we obtain a reasonable scaling behavior.

Figure \ref{dvsmu} shows the measured fractal dimensions $D$ as a function of
$\mu$. We computed $D$ by averaging over the dimensions for each individual
cluster, obtained from the corresponding scaling plot of $R_G$ against $M$. The
error bars were obtained from the statistical fluctuations.  Quantitatively the
same result is obtained when we compute $D$ from the slopes of the
geometrically averaged $R_G$ vs $M$ plots (Fig. \ref{mvsrg}). Also shown in
Fig. \ref{dvsmu} is the prediction of the scaling argument in Eq. (\ref{D2}).
Good agreement with our theory is observed far away from the transition points
for $\mu\simeq 1$ and $\mu\agt 2.5$. At the transition points $\mu={D_{DLA}}$
and 2,
the expected discrepancies due to the logarithmic corrections in Eq.
(\ref{logcorr}) are observed.

%The $\mu={D_{DLA}}$ point is the transition between the regimes dominated by
%%small and
%huge particles. Because of the logarithmic corrections, the sharp transition
%cannot be observed numerically. In fact, in earlier simulations of surface
%growth with power-law noise, the analogous transition was not identified as
%well.  Recently, Lam and Sander reported direct numerical evidence for the
%existence of the logarithmic correction in surface growth.  This provides
%strong support for their exact scaling theory, which we adapted here.  In the
%current case, limited by the rather small cluster size we used and the strong
%statistical fluctuations, we cannot study this correction quantitatively.

In summary, we propose a variant of $DLA$ in which the random walkers have
random radii of very broad power law distribution. A scaling theory predicts
that, as the distribution becomes narrower, there exists a transition between
regimes dominated by large and small particles. The fractal dimension can be
calculated exactly for the former regime. At the transition point we find a
logarithmic correction to scaling. An analogous scaling theory leading to a
similar transition has been verified numerically for the case of surface growth
with power law noise. We generated $1200$ clusters of our $DLA$ variant. When
huge particles dominate, the morphology of the cluster is different from that
of $DLA$ and might be related to that of the cluster-cluster aggregation model.
The measured fractal dimension is in reasonable agreement with the scaling
theory.  The numerical precision is limited by strong statistical fluctuations
and finite size effects, which manifest themselves as logarithmic corrections
at the transition points.

We would like to thank H.~J.~Herrmann, J.~Lee, M.~Leibig,
H.~E.~Stanley and A.~Vespignani for discussions. LMS is supported by NSF Grant
DMR 91-17249.

\begin{figure}[htb]
\caption{\label{clusters}
Typical clusters at (a) the small particles dominating regime ($\mu = 2.5$),
(b) the transition point ($\mu=1.713$), and (c) the large particles
dominating regime ($\mu=1.3$).}
\end{figure}

\begin{figure}[htb]
\caption{\label{mvsrg}
Dependence of the cluster mass $M$ on the radius of gyration $R_G$ for three
selected values of $\mu$. The geometrical mean is adopted for the ensemble
averaging.}
\end{figure}

\begin{figure}[htb]
\caption{\label{dvsmu}
Dependence of the fractal dimension $D$ on the exponent $\mu$. The
solid line denotes our numerical results, whereas the dashed line
marks the prediction of the scaling theory. The systematic
discrepancies at the transition points $\mu\simeq 1.713$ and $\mu=2$
are due to finite size effects in the form of logarithmic corrections.
}
\end{figure}

\end{document}